\documentclass[prl,twocolumn,showpacs,10pt]{revtex4-1} 
\usepackage{amsmath,amssymb,graphicx,bm}

\usepackage{color,times,units}

\definecolor{darkblue}{rgb}{0, 0, 0.8}
\usepackage[colorlinks=true, breaklinks=true, linkcolor=darkblue, citecolor=darkblue, urlcolor=darkblue]{hyperref}
\newcommand{\ket}[1]{\left| #1\right\rangle}

\definecolor{rkcol}{rgb}{0.8, 0, 0.5}

\begin{document}
	
\title{Three-dimensional trapping of individual Rydberg atoms in ponderomotive bottle beam traps}
	
\author{D. Barredo}
\author{V. Lienhard}
\author{P. Scholl}
\author{S. de L\'es\'eleuc}
\altaffiliation{Present address: Institute for Molecular Science, National Institutes of Natural Sciences, Myodaiji, Okazaki 444-8585, Japan.}
\author{T. Boulier}
\author{A. Browaeys}  
\author{T. Lahaye}
\affiliation{Laboratoire Charles Fabry, Institut d'Optique Graduate School, CNRS, Universit\'e Paris-Saclay, 91127 Palaiseau Cedex, France}	
	
\begin{abstract}
We demonstrate three-dimensional trapping of individual Rydberg atoms in holographic optical bottle beam traps. Starting with cold, ground-state $^{87}$Rb atoms held in standard optical tweezers, we excite them to $nS_{1/2}$, $nP_{1/2}$, or $nD_{3/2}$ Rydberg states and transfer them to a hollow trap at 850 nm. For principal quantum numbers $60 \leqslant n \leqslant 90$, the measured trapping time  coincides with the Rydberg state lifetime in a 300~K environment. We show that these traps are compatible with quantum information and simulation tasks by performing single qubit microwave Rabi flopping, as well as by measuring the interaction-induced, coherent spin-exchange dynamics between two trapped Rydberg atoms separated by 40 $\mu$m. These results will find applications in the realization of high-fidelity quantum simulations and quantum logic operations with Rydberg atoms.
\end{abstract}
	
\maketitle
	
Neutral atoms excited to Rydberg states are an attractive platform for large-scale quantum simulation and computation~\cite{Jaksch2000,Saffman2010}. The strong, controllable interactions between these excited states can be used to implement high-fidelity quantum logic gates, or to engineer various types of spin Hamiltonians that are difficult to study on classical computers~\cite{Weimer2010}. These ideas have been intensively explored in the last years and several important milestones have been achieved~\cite{Saffman2016}. Prominent examples of this progress are the demonstration of strong optical non-linearities~\cite{Pritchard2010}, single-photon sources~\cite{Dudin2012}, conditional phase shifters~\cite{Firstenberg2013}, single-photon transistors~\cite{Tiarks2014,Gorniaczyk2014}, the experimental realizations of two-qubit gates~\cite{Wilk2010,Isenhower2010,Biedermann2016,Levine2018}, or the first quantum simulations of spin models with tens of particles in optical lattices~\cite{Schauss2012,Schauss2015,Zeiher2017} and in arrays of optical tweezers~\cite{Labuhn2016,Bernien2017,Kim2018,deLeseleuc2019}. 
	
In none of the above experiments were the Rydberg atoms trapped. However, control over the motion of Rydberg atoms during gate operation and in quantum simulations is advantageous, since finite atom temperatures and mechanical forces due to the strong interactions between the particles ultimately limit quantum state fidelities~\cite{Saffman2005,Saffman2016} and the available time for coherent dynamics~\cite{Barredo2015,Bernien2017, deLeseleuc2018_PRA}. Rydberg trapping is also a prerequisite for precision measurements of fundamental constants using circular Rydberg states~\cite{Ramos2017,Jentschura2008} or positronium~\cite{Cassidy2018}.  
	
\begin{figure}[b]
\centering
\includegraphics[scale=0.95]{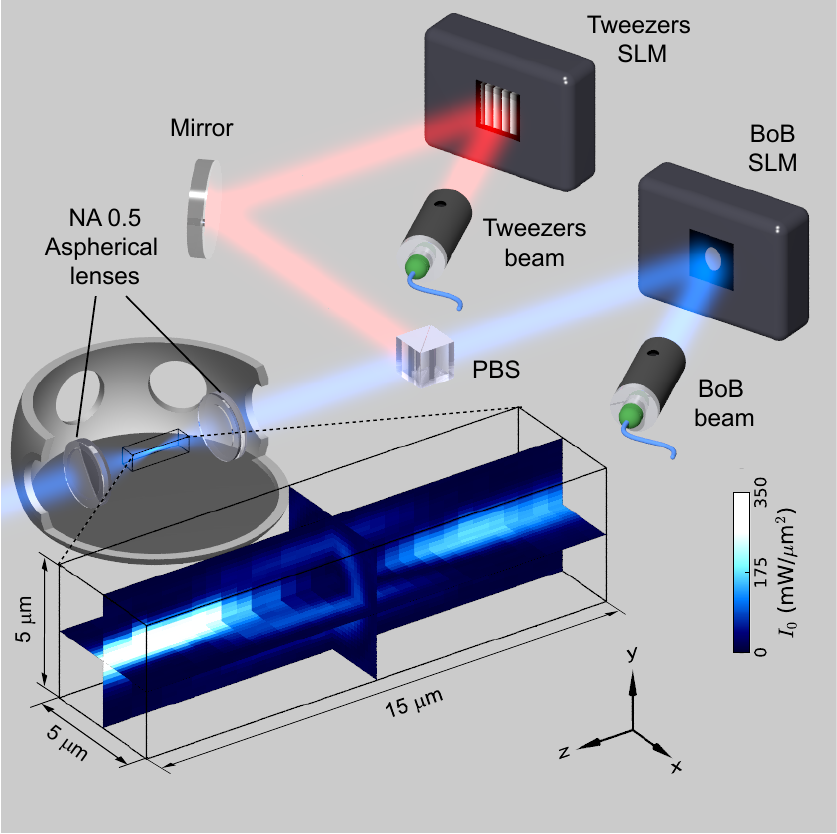}
\caption{Main components of the experimental setup. We use a spatial light modulator (SLM) and a high NA aspherical lens  to generate a BoB trap for Rydberg atoms (blue beam, see text). The zoom shows two-dimensional cuts of the reconstructed light intensity distribution near the focal plane, measured using an electrically-tunable lens (not shown) to scan the images along the optical axis. To load the BoB trap, we use regular optical tweezers (red beam) created by light reflected from a second SLM and superimposed onto the BoB trap beam using a polarizing beam splitter (PBS).}
\label{fig:fig1}
\end{figure}	
	
To date, clouds of Rydberg atoms have been confined to three-dimensional, millimeter-size regions using static magnetic~\cite{Choi2005,Anderson2013,Bounds2018} or electric fields~\cite{Vliegen2007, Hogan2008}. In an inhomogeneous AC electric field that oscillates faster than any internal frequency of the Rydberg atom, the weakly-bound Rydberg electron experiences an oscillating force that can be used for laser trapping of the Rydberg atom~\cite{Anderson2011}. The so-called ponderomotive potential, which is the time-averaged kinetic energy of the nearly free Rydberg electron oscillating in the laser field, is proportional to the light intensity. Therefore, to obtain a 3D trap, one must create a dark region surrounded by light in all directions; since the atom trapping arises mainly from the ponderomotive potential experienced by the electron, such traps can be used to confine Rydberg states whatever their $n,l,j,m_j$ quantum numbers. 

Rydberg atoms have been efficiently confined in the tight potentials of ponderomotive optical lattices~\cite{Anderson2011, Li2013}, but so far only in one dimension. Here we go beyond those initial demonstrations to show efficient three-dimensional trapping of cold individual Rydberg atoms in micron-size optical potentials~\cite{Zhang2011,Zhang2012_thesis}. We use holography to create bottle beam (BoB) traps~\cite{Ozeri1999,Xu2010} which are deterministically loaded with single Rydberg atoms. We  characterize the depth and trapping frequencies of these traps and observe that the trapping time for principal quantum numbers in the range $60 \leqslant n \leqslant 90$ is mainly limited by the lifetime of the Rydberg states in the presence of blackbody radiation at 300 K. Finally, we illustrate the compatibility of these traps with quantum simulation by driving Rabi flopping between different Rydberg states, and by observing the coherent exchange of internal states induced by the dipole-dipole interaction for two atoms confined in BoB traps separated by 40 $\mu$m. 
	
The experimental setup is based on the one described in Ref.~\cite{Barredo2018}. Briefly, we use a spatial light modulator (SLM) to create arbitrary arrays of micron-size optical tweezers (Fig.~\ref{fig:fig1}). To generate the BoB traps that will host Rydberg atoms, we follow the procedure of Refs.~\cite{Chaloupka1997,Xu2010}. We use a second SLM to imprint on another light beam at 850 nm a phase pattern that is composed of two terms added modulo $2\pi$: (i) the standard hologram to create the desired number and positions of point traps, and (ii) a centered disk of radius $r_0$ where the phase is shifted by $\pi$. The value of $r_0$ is adjusted such that, for each trap, destructive interference occurs between the central and outer parts of the Gaussian beam at the focal point, thus creating the needed dark region surrounded by light in all directions. The local maximum of light intensity encountered when moving away from the origin is smallest ($\sim 10\%$ of the maximal intensity) on two ``escape cones'' originating from the trap center. By changing the diameter of the beam impinging on the SLM and adjusting $r_0$ accordingly, the size of the trapping region can be tuned.  An example of BoB trap created using this method is shown in the inset of Fig.~\ref{fig:fig1}, where we used an electrically-tunable lens in the imaging path to record the 3D light intensity distribution near the focal plane of the aspherical lens. To allow for an efficient transfer of the atoms between the two different traps (for ground-state and Rydberg atoms), the light from the BoB SLM is then overlapped with the standard optical tweezers beam using a polarizing beam splitter (PBS), ensuring that both traps are centered at the same position. 
	
Far from resonances and  in a Born-Oppenheimer-like approximation, the Rydberg atom trapping potential in a light field of intensity $I$ and angular frequency $\omega_L$ can be described as~\cite{Dutta2000}:
\begin{equation}
U_{nljm_j}(\boldsymbol{R})=\int d^3 r \, V_P(\boldsymbol{R+r}) \, |\psi_{nljm_j}(\boldsymbol{r})|^2 \, .
\label{eq:ponderomotive_potential}
\end{equation} 
Here, $V_P(\boldsymbol{r})=e^2 I(\boldsymbol{r})/{(2 m_e \epsilon_0 c \omega_L^2)}$ is the repulsive ponderomotive shift experienced by the nearly free electron of mass $m_e$ and charge $e$, $\psi$ is the Rydberg wave function, $\boldsymbol{R}$ is the center of mass coordinate of the atom, and $\boldsymbol{r}$ is the relative position of the Rydberg electron. 

\begin{figure}[t]
\centering	\includegraphics[scale=1.0]{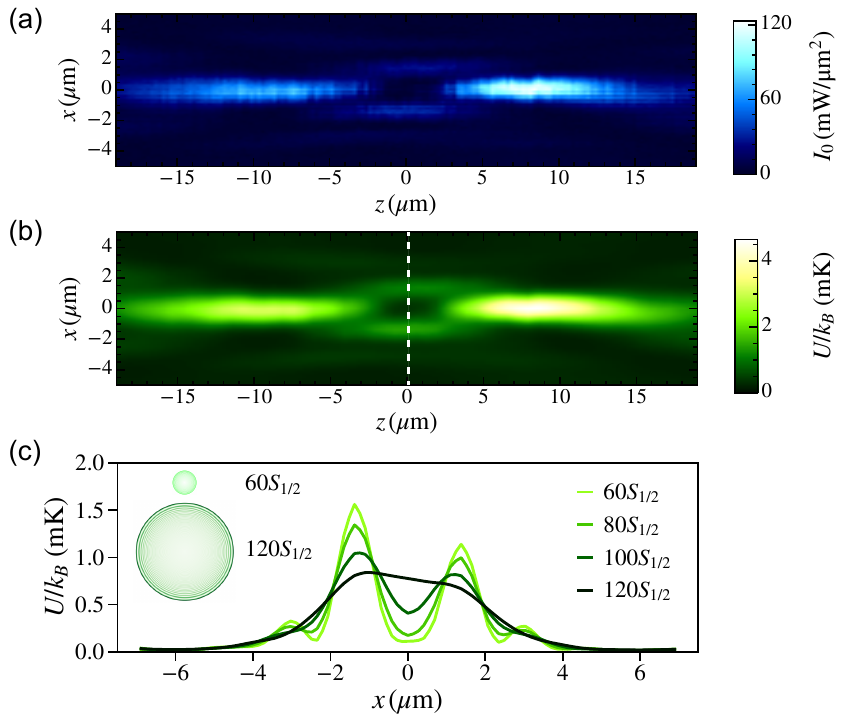}
\caption{(a) Example of measured light intensity distribution near the focal plane of the aspherical lens for a total beam power of 400 mW. (b) Trapping potential for the $84 S_{1/2}$ Rydberg state calculated from Eq.~(\ref{eq:ponderomotive_potential}) and the intensity distribution shown in (a). (c) One-dimensional cuts of the trapping potential along $z=0$ (dashed line in (b)) for different $nS_{1/2}$ Rydberg states. A schematic of the $60S$ and $120S$ Rydberg orbitals (to scale) is shown as an inset.}
\label{fig:fig2}
\end{figure}

An example of such calculation for the $84S_{1/2}$ Rydberg state is shown in the axial cut displayed in Fig.~\ref{fig:fig2}b, calculated from the measured light intensity displayed on Fig.~\ref{fig:fig2}a. The trapping potential is harmonic in the axial direction $z$, whereas in the transverse directions it has approximately a quartic form close to the center, and maxima separated by $\sim2.5 \,\mu$m. From the measured three-dimensional light intensity distribution and a total power of 400~mW (Fig.~\ref{fig:fig2}a), we obtain minimum trap barriers (saddle point along the escape cones) of around 0.6~mK (Fig.~\ref{fig:fig2}b). When the variation of the field starts to be substantial over a length scale comparable to the atom size, the trapping potential $U_{nljm_j}(\boldsymbol{R})$ depends strongly on the specific Rydberg state. This is illustrated in the one-dimensional cuts along the transverse direction for different Rydberg states displayed in Fig.~\ref{fig:fig2}c.  When the orbital radius $\sim n^2 a_0$ of the atom becomes comparable to the trap size (here, for $n\sim 120$), the potential does not have a local minimum any longer, and Rydberg atoms are then repelled from the BoB trap. 
	
\begin{figure}[t]
\centering
\includegraphics[scale=1.0]{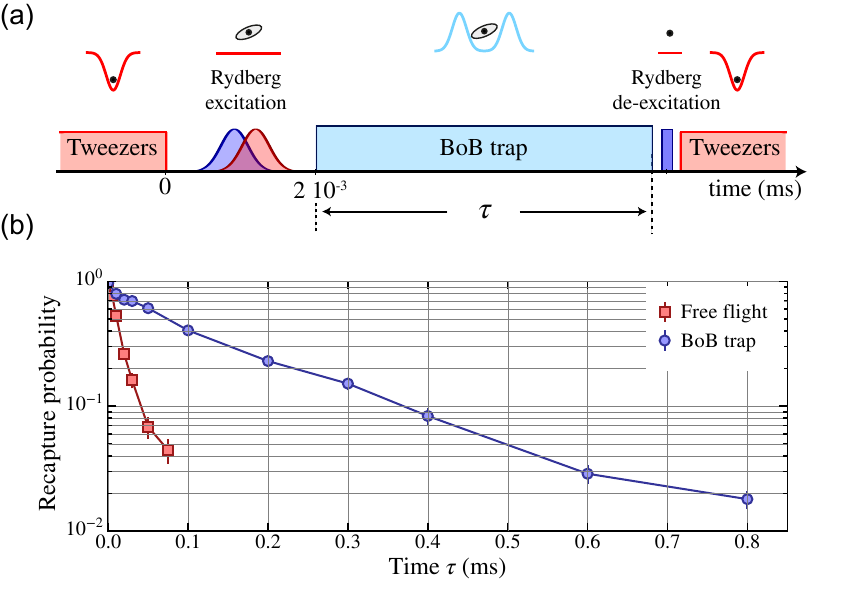}
\caption{(a) Time sequence of the experiment. Ground-state atoms are initially loaded in optical tweezers. The tweezers are then switched off for Rydberg excitation. After 2 $\mu$s the BoB trap is applied for a variable duration $\tau$. Following Rydberg de-excitation in free-flight, the tweezers are switched on again and the atoms are imaged. As the BoB trap is strongly repulsive for ground state atoms, only atoms that were in the Rydberg state before de-excitation are recaptured. (b) Recapture probability for atoms excited to the $84S_{1/2}$ state as a function of the trapping time $\tau$ (disks). The recapture curve in the absence of BoB trap (squares) is shown for comparison. Error bars are the standard error of the mean.}
\label{fig:fig3}
\end{figure}	
	
The experimental sequence is shown in Fig.~\ref{fig:fig3}. First, $^{87}$Rb atoms from a magneto-optical trap (MOT) are loaded in the optical tweezers (with temperatures in the range $3-130\,\mu$K depending on the experiment). We detect the occupancy of each trap by collecting the fluorescence of the atoms at 780 nm with an electron-multiplying CCD camera in 20 ms. After successful detection of a filled trap, atoms are optically pumped in the ground state $\ket{g}=\ket{5S_{1/2},F=2,m_F=2}$, the optical tweezers are switched off, and a two-photon STImulated Raman Adiabatic Passage (STIRAP) prepares the target Rydberg state $\ket{r}$ with $\sim 90 \%$ efficiency. We then switch on the BoB trap for a variable duration $\tau$. Finally, a Rydberg de-excitation pulse, resonant with the $\ket{r} - 5P_{1/2}$ transition, brings the population in $\ket{r}$ back to the ground state $\ket{g}$ via spontaneous decay from $5P_{1/2}$, and the atom is then imaged by fluorescence. The result of one of these experiments is plotted in Fig.~\ref{fig:fig3}. Without the BoB trap (red curve), this release and recapture experiment can be used to measure single atom temperatures~\cite{Tuchendler2008}. The recapture probability for atoms in $84 S_{1/2}$ is reduced by a factor ten after $30\, \mu$s. From this curve we extract a single atom temperature of $\sim 130 \, \mu$K. When the BoB trap is applied (blue curve), the recapture probability is drastically enhanced, and only 30~\% of the atoms are lost in the same amount of time, consistent with the expected Rydberg lifetime (see below). We do not observe any appreciable heating of the atoms after they are transferred back to the ground-state tweezers.
	
Following a procedure similar to the one in Ref.~\cite{Sortais2007}, we determined the typical radial oscillation frequency of the Rydberg atom in the trap. After the atoms are loaded into the BoB trap, we excite breathing oscillations by switching off the trap for three microseconds. During this trap-off time the atom leaves the bottom of the potential. We then switch on the trap again for a variable time $T$, during which the atom oscillates in the trap with a typical frequency $\omega_{\rm BoB}$. The recapture probability after a fixed switch off time oscillates with a frequency $2 \omega_{\rm BoB}$. For atoms excited to the $84S_{1/2}$ state we observe oscillations in the recapture probability that are well reproduced by Monte-Carlo simulations taking into account the experimental parameters, where the sampled atom trajectories inside the trapping field are computed classically~\cite{SM}. For a trap depth of $\sim 0.6$~mK the atoms oscillate in the traps with a measured frequency of $\omega_{\rm BoB} / (2\pi) = 59\pm3$~kHz, in fair agreement with the simulations, which predict $59$~kHz for the same parameters. The necessary power for efficient trapping depends on the trap size, the initial temperature of the atoms, and the target Rydberg state. For our smallest trap (with maxima separated by $\sim 2 \, \mu$m in the radial direction), and single atom temperatures of $\sim 3 \,\mu$K, we need only 20~mW of power to reach our higher trapping efficiencies for the $60S_{1/2}$ Rydberg state, which shows that it will be possible to scale  this technique up to large arrays of BoB traps.
	
\begin{figure}[t]
\centering
\includegraphics[scale=1.0]{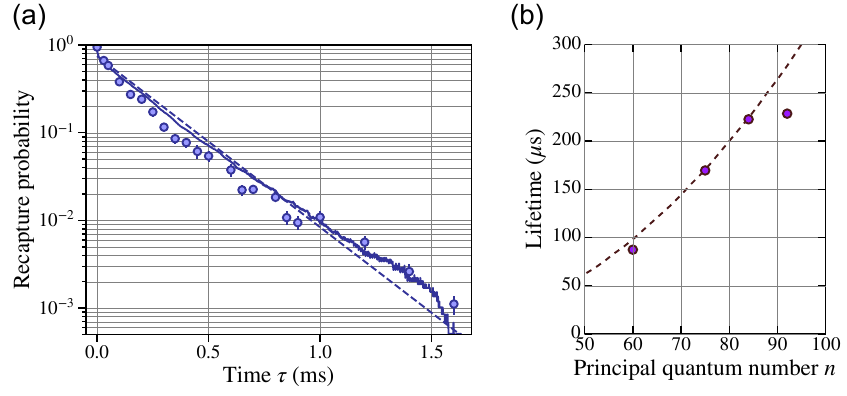}
\caption{(a) Recapture probability of a $84S_{1/2}$ Rydberg atom, as a function of the time spent in the BoB trap, showing a roughly exponential decay with lifetime $222\pm3\,\mu$s (dashed line). The solid line is the result of a simulation without any adjustable parameter (see text). (b) Measured lifetimes of various $nS_{1/2}$ states; the dashed line is the theoretical value~\cite{Beterov2009}. }
\label{fig:fig4}
\end{figure}

To evaluate the quality of the BoB potential as a trap for Rydberg atoms, we measured the trap-decay times for different Rydberg states. The recapture probability as a function of the time spent in the BoB trap is shown for the $84S_{1/2}$ state in Figure~\ref{fig:fig4}a. An exponential fit to the experimental data (dashed line) gives a $1/e$ decay time of $222\pm3\;\mu$s, in excellent agreement with the calculated lifetime $228\;\mu$s of this state at 300~K~\cite{Beterov2009,Sibalic2017,Archimi2019}. The solid line is the result of a Monte-Carlo simulation of the atomic trajectories in the BoB trap potential (calculated using Eq.~(\ref{eq:ponderomotive_potential}) and the measured light intensity distribution) and taking into account not only radiative decay, but also the possibility for some energetic atoms to spill over the trap barrier. All the parameters entering the simulation were given their experimentally measured values~\cite{SM}.  Figure~\ref{fig:fig4}b shows the results of the same experiment for other $nS_{1/2}$ states; the measured decay times match almost perfectly the expected lifetime (dashed line) up to $n\sim90$. This indicates that atom losses are mainly due to Rydberg decay, and that other processes, such as photoionization~\cite{Saffman2005,Potvliege2006,Zhang2011}, are negligible. We also performed the same measurement for the $84P_{1/2}$ and $82D_{3/2}$ states (not shown) and obtained lifetimes $\sim20\%$ shorter than their theoretical values. 

Beyond $n = 90$, the extension of the radial wave function becomes comparable to the trap size and trapping is less efficient. As a consequence, some atoms can escape the trap, and the loss rate is significantly higher than expected from the Rydberg lifetime alone. Beyond $n \sim 100$ the potential does not have a local minimum and the Rydberg atoms are quickly expelled from the BoB traps. Monte-Carlo simulations~\cite{SM} show that for $n=60$, atoms can be lost only through the narrow escape cones of the BoB trap, while for $n=100$, we observe much more pronounced losses over the radial barrier due to the small depth of the trap.

\begin{figure}[t]
\centering
\includegraphics[scale=1.0]{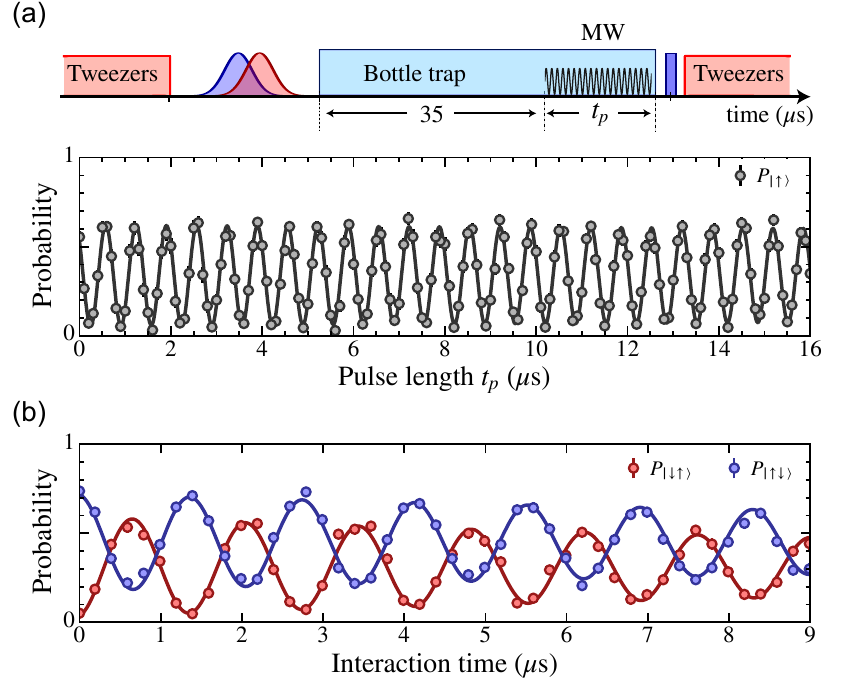}
\caption{(a) Rabi oscillation between the states $\ket{\uparrow} = \ket{82 D_{3/2}, m_j=3/2}$ and $\ket{\downarrow} = \ket{83 P_{1/2}, m_j=1/2}$. The total trapping time is $50\,\mu$s. (b)~Excitation-hopping oscillations in the population of the pair states $\ket{\uparrow\downarrow}$ and $\ket{\downarrow\uparrow}$, mediated by the dipole-dipole interaction between the Rydberg states $\ket{\uparrow}$ and $\ket{\downarrow}$ at a distance of $40\,\mu$m. The temperature of the atoms is $\sim 3\, \mu$K. Solid lines are damped sine fits to the data. Error bars represent the standard error of the mean and are most often smaller than the symbol size.}
\label{fig:fig5}
\end{figure}

We now investigate the performance of our BoB trap for quantum simulation tasks by studying its compatibility with one- and two-qubit operations. For low principal quantum numbers ($n<60$), the trapping potential corresponds approximately to the one of a free electron, and it is therefore almost independent of $n$ or the atom's angular momenta. For higher Rydberg states the wave function of the electron must be considered in the convolution (\ref{eq:ponderomotive_potential}). Most quantum simulation experiments, however, involve only couplings to adjacent Rydberg levels, for which changes in the squared radial part of the wave function play only a minor role in the effective potential. The angular part, in turn, is identical for states with different quantum number $l$, but same $j$, $m_j$ (e.g., for $\ket{nS_{1/2}, m_j=1/2}$ and $\ket{nP_{1/2}, m_j=1/2}$). This `quasi-magic' trapping condition strongly suppresses  differential light shifts in Rabi oscillations between different trapped Rydberg states. In a first experiment, we analyze the coherence in the spin manipulation of a single trapped atom. We apply the BoB trap for a total time of $50\, \mu$s. We let the atoms move in the trapping potential during $\sim 35\, \mu$s and then we drive microwave Rabi oscillations between the states $\ket{\uparrow} = \ket{82 D_{3/2}, m_j=3/2}$ and $\ket{\downarrow} = \ket{83 P_{1/2}, m_j=1/2}$, while the atoms are still confined (Fig.~\ref{fig:fig5}a). We observe spin flip oscillations without appreciable damping, despite the two states having different angular wave functions. The constant finite contrast of the oscillations is mainly due to the excitation efficiency and the limited lifetime of the involved Rydberg states. We measured similar Rabi flopping curves even for atom temperatures as high as $130\, \mu$K, where the atoms explore a large volume of the BoB trap.
	
As a final illustration of the usefulness of our trapping scheme for quantum simulation, we measure spin-exchange dynamics driven by dipole-dipole interaction between two atoms. We create two traps separated by a distance of $R=40\,\mu$m (Fig.~\ref{fig:fig5}b). Immediately after loading the Rydberg atoms in the BoB traps, we use a resonant microwave field and local addressing~\cite{deLeseleuc2017} to prepare the first atom in $\ket{\uparrow}$ and the second one in $\ket{\downarrow}$. In these two Rydberg states, the atoms are resonantly coupled by a dipole–-dipole interaction of strength $U_{\rm dd}/h =  C_3/ R^3 \sim 0.4$ MHz, and the dynamics is governed by an XY-spin Hamiltonian $H=U_{\rm dd}(\sigma_1^+\sigma_2^- + \sigma_1^-\sigma_2^+$), where $\sigma_i^\pm$ are the Pauli matrices for atom $i={1,2}$. We observe coherent spin exchange between the pair states $\ket{\uparrow \downarrow}$ and $\ket{\downarrow \uparrow}$. This experiment demonstrates the compatibility of our Rydberg trapping scheme with quantum simulations using current experimental setups, with levels of contrast and damping comparable or better that previously achieved~\cite{Barredo2015,deLeseleuc2017}.

In conclusion, we have demonstrated laser trapping of individual Rydberg atoms in microscopic potentials and shown the suitability of such traps for quantum information tasks. This trapping scheme can be extended to larger arrays with moderate laser power for atoms which can be cooled to the motional ground state before Rydberg excitation~\cite{Kaufman2012,Thompson2013}. In addition, the ponderomotive traps can, in principle, be used for circular Rydberg atoms in a cryogenic environment, opening the door to unprecedented trapping times~\cite{Nguyen2018}. The results presented here are suited for experiments that occur entirely in the Rydberg manifold. For quantum logic gates or for quantum simulation involving also the ground state in addition to Rydberg states, it would be possible, by using shorter wavelength trapping light, to realize BoB traps that satisfy the ground-Rydberg `magic' condition and would minimize heating and decoherence rates~\cite{Zhang2011,Piotrowicz2013}. 
	
{\it Note:} Two-dimensional trapping of circular Rydberg atoms in the ponderomotive potential of a hollow laser beam has been observed recently at Coll\`ege de France~\cite{PrivateComm}.	
	
\begin{acknowledgements}
This work benefited from financial support by the EU (FET-Flag 817482, PASQUANS), by the R\'egion \^Ile-de-France in the framework of DIM SIRTEQ (project CARAQUES), and by the IXCORE-Fondation pour la Recherche. T.~B. acknowledges the support of the European Marie Sk{\l}odowska-Curie Actions (H2020-MSCA-IF-2015 Grant No. 701034).
\end{acknowledgements}
	
\bibliography{Refs}
	
\end{document}


\title{Supplemental Material: Three-dimensional trapping of individual Rydberg atoms \\in ponderomotive bottle beam traps}

\author{D. Barredo, V. Lienhard, P. Scholl, S. de L\'es\'eleuc, T. Boulier, A. Browaeys and  T. Lahaye}
\affiliation{Laboratoire Charles Fabry, Institut d'Optique Graduate School, CNRS, Universit\'e Paris-Saclay, 91127 Palaiseau Cedex, France}

\maketitle

\section{Numerical simulation of the atomic motion in the BoB trap}

To assess the quality of the BoB traps, we performed classical Monte-Carlo simulations of the atomic motion in the trapping potential. This allowed us to compute the recapture probability obtained in a release and recapture experiment (see Fig. 3a of the main text for the sequence) and compare it with experimental data. We simulate the full experimental sequence, including the two free-flight stages during Rydberg excitation and de-excitation, and the atomic motion in the BoB trap for a time $\tau$. To determine if an atom is recaptured at the end of the sequence, we compare its kinetic energy to the potential energy of the ground state trap at the final position of the atom. 

We account for the finite Rydberg lifetime at 300~K by introducing, for each Rydberg state, a single exponential probability distribution function in the dynamics. This approximation is justified because the two processes contributing to the Rydberg lifetime (decay to low-lying states and blackbody radiation-induced emission to other Rydberg states) produce the same outcome in this experiment: a loss of the atom in the final fluorescence image. Atoms that are not successfully excited by the STIRAP pulses, or Rydberg atoms that spontaneously decay to the ground state, are quickly repelled from the trapping region by the BoB potential and not recaptured at the end of the sequence. On the contrary, Rydberg atoms transferred to other Rydberg states by black-body radiation can still remain trapped in the BoB, but are not projected to the ground state by the de-excitation pulse. As a consequence, they are repelled by the repulsive potential of the standard optical tweezers before imaging, and finally lost.

For each simulation we use the same parameters as in the experiment (namely, a total power of 400 mW, the recorded three-dimensional trap intensity distribution shown in Fig. 2a of the main text, and the measured STIRAP excitation efficiency). The final result of the simulation is obtained by averaging over $\sim10^4$ realizations, starting from random atom positions (with rms values $\sigma_r = \sqrt{k_B T/m \omega_r^2}\sim 230\, \rm{nm}$; $\sigma_z = \sqrt{k_B T/m \omega_z^2}\sim 1.15\,\mu\rm{m}$) and velocities (rms value $\sigma_v = \sqrt{k_B T/m}\sim 110\, \rm{nm}/ \mu\rm{s}$),  according to a thermal atomic distribution in the standard optical tweezers at $T=130\,\mu$K, with radial and longitudinal trapping frequencies $\omega_r / (2\pi) = 75\, \rm{kHz}$, and $\omega_z / (2\pi) = 15\, \rm{kHz}$, respectively.

\begin{figure*}[b]
\centering
\includegraphics[width=\textwidth]{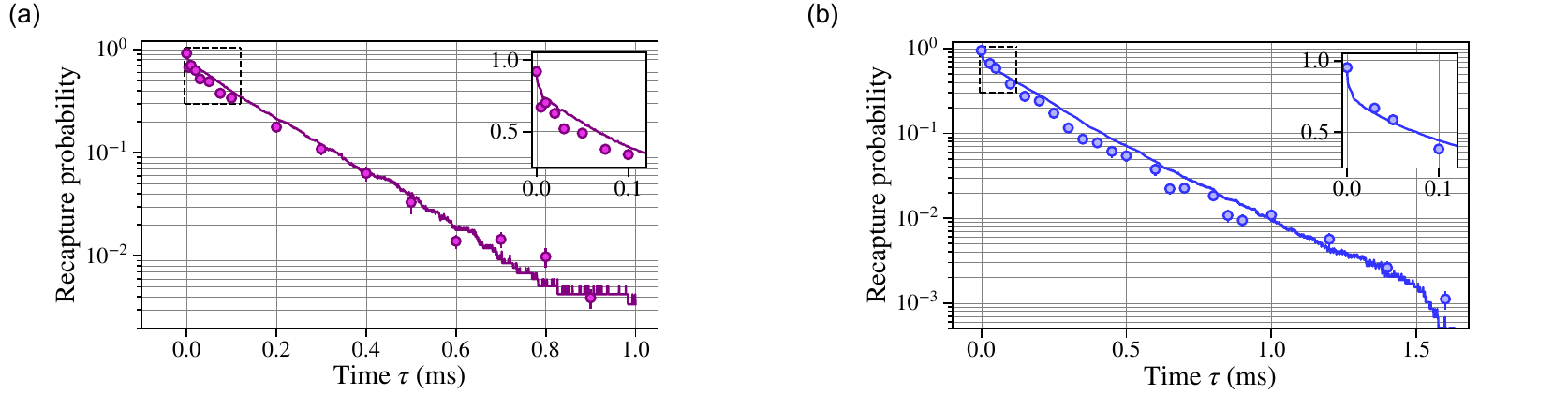}
\caption{Simulated recapture probability as a function of the trapping time $\tau$ (solid lines) for the Rydberg states $75S_{1/2}$ (a) and $84S_{1/2}$ (b). The experimental data (disks) are shown for comparison. Insets are zooms on the short-time behavior (dashed lines) showing, in linear scale, a fast atom loss at short times.}
\label{fig:S1}
\end{figure*}

Figure S\ref{fig:S1} shows the results of the simulation for two different Rydberg states, $75S_{1/2}$, and $84S_{1/2}$. The agreement between the experimental data and the simulation with no free parameters is very good over the full time span. At short times ($\tau < 50 \,\mu\rm{s}$) we observe a fast decrease in the recapture probability (see inset) both in the experimental data and in the simulation. In the simulation, this initial drop has two contributions: (i) about $10-15\%$ of this decrease (depending on the dataset) corresponds to the loss of atoms that remain in the ground state after the STIRAP excitation pulses; (ii) the other $10\%$ is due to energetic atoms having enough kinetic energy to overcome the potential barrier of the BoB trap. 

\section{Analysis of the mechanical losses in the BoB trap}

\begin{figure*}[t]
	\centering
	\includegraphics[width=17cm]{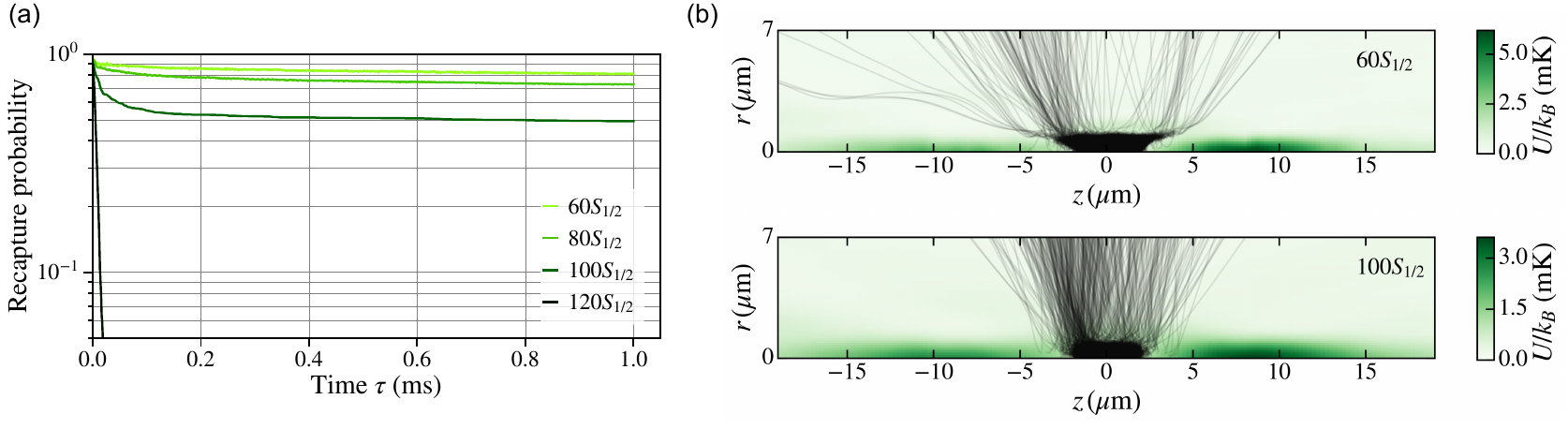}
	\caption{(a) Simulated recapture probability as a function of the trapping time for different Rydberg states assuming an infinite Rydberg lifetime. (b) Classical trajectories of Rydberg atoms evolving in the BoB potential. For the $60S_{1/2}$ state mechanical losses occur close to the saddle points (`escape cones'), whereas for $100S_{1/2}$ atoms leave the trap uniformly in the radial direction.}
	\label{fig:S2}
\end{figure*}

In order to investigate the (small) initial  atom loss and to distinguish ``mechanical losses'' due to inefficient or unstable trapping from atom losses due to Rydberg state decay, we performed the same simulation as described in the previous section, but now assuming an infinite Rydberg lifetime. The results are shown in Fig. S\ref{fig:S2}a.

For $n\le 100$, we observe again the fast initial decrease in the recapture probability in a time scale of $\sim 50\, \mu\rm{s}$, before reaching an almost stationary value at around $\tau = 1$~ms. For times longer than $\tau\sim 50\,\mu\rm{s}$ the decay rate due to mechanical losses is comparatively much lower than the inverse of the Rydberg state lifetime. The fact that extra losses of atoms occur mainly at short times allows us to measure Rydberg trapping times close to the theoretical Rydberg state lifetimes. For $n=120$, the potential does not have a local minimum (see Fig. 2 of the main text), and Rydberg atoms are quickly repelled from the trapping region.

The atom trajectories shown in Fig. S\ref{fig:S2}b give us information about the escape regions in the trapping potential. For the $60S_{1/2}$ atoms escape mainly though the saddle points. For higher Rydberg states, the effect of the convolution is to smooth the saddle cones at the expense of decreasing the effective potential barrier, and atoms escape the trap in the radial direction in a more homogeneous manner.

\section{Oscillation frequency of the atoms in the BoB trap}

\begin{figure*}[b]
\centering
\includegraphics[width=17cm]{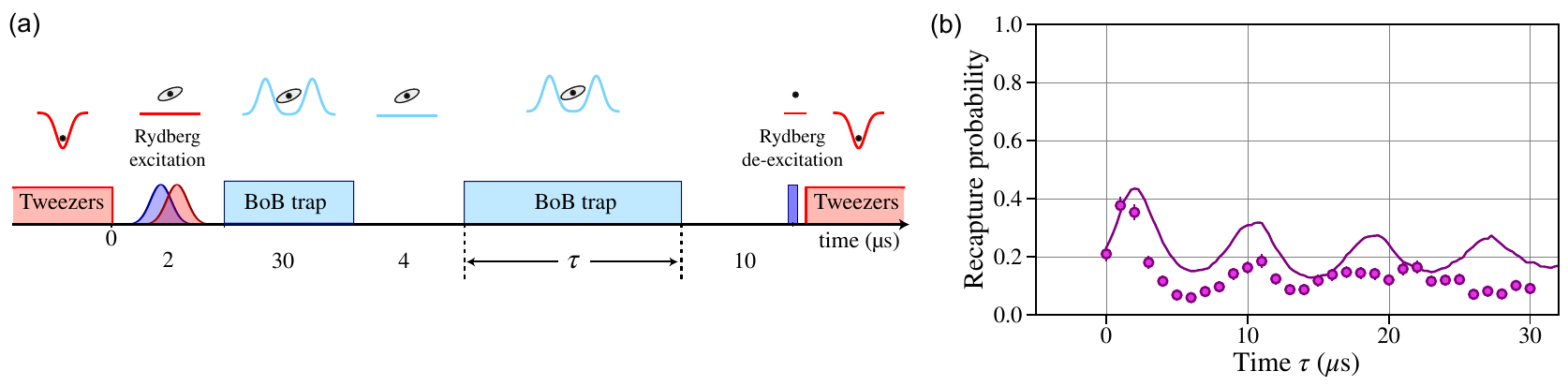}
\caption{(a) Sketch of the experimental sequence to measure the oscillation frequency of the atoms in the BoB trap. (b) Experimental data (disks) and classical simulation (solid curve) of the 'release and recapture' experiment performed using the sequence in (a) for a trap depth of 0.6 mK.}
\label{fig:S3}
\end{figure*}

We present in this section an experiment to measure the typical radial  trapping frequency in the BoB trap. The experimental sequence is shown in Fig. S\ref{fig:S3}a. We first transfer the Rydberg atom to the BoB trap. We then wait for $30\,\mu\rm{s}$. During this preliminary waiting time, the hottest atoms escape, as explained in the previous section. Then, we follow the same approach as in \cite{Tuchendler2008}, replacing the standard optical tweezers by our BoB trap. We switch off the trap during $4\,\mu\rm{s}$ to excite the radial breathing mode. Then, atoms oscillate in the BoB trap for a varying duration $\tau$. We release the atoms for $10\,\mu\rm{s}$ and finally de-excite and recapture them in the standard optical tweezers. For a harmonic trap, the recapture probability is expected to oscillate at $2\omega$, with $\omega$ the trapping frequency.

Fig. S\ref{fig:S3}b shows the measured recapture probability for a trap depth of 0.6 mK. From the observed oscillations we extract a trapping frequency, $\omega_{\rm{BoB}} / (2\pi) = 59\pm 3\,\rm{kHz}$. The observed damping of the oscillations is most likely due to the finite atom temperature and the anharmonicity of the trapping potential. In the simulation, we follow the same procedure described above to numerically compute the recapture probability with no adjustable parameters. The result (solid line) is in reasonable agreement with the experimental data, as we find a trapping frequency $\omega_{\rm{BoB}} = 2\pi\times 59\,\rm{kHz}$.